\documentclass[10pt,letterpaper,twocolumn]{article}
\usepackage{ol2}

\usepackage{amssymb}
\newcommand{\figwidth}{3.2in}

\renewenvironment{abstract}
{\noindent\begin{center}{\footnotesize Submitted to OSA Dec 23, 2008
} \vskip4pt
\begin{minipage}{34.25pc} \parindent.2in
   \noindent \footnotesize \rm}{\hskip.07in  \\ \hfil \end{minipage}\end{center}}

\begin{document}

\twocolumn[

\title{Accurate calculation of the local density of optical states in inverse-opal photonic crystals}

\author{Ivan S. Nikolaev$^{1}$, Willem L. Vos$^{1,2}$, and A. Femius Koenderink$^{1,*}$}
\address{$^1$Center for Nanophotonics, FOM Institute for Atomic and Molecular Physics
(AMOLF), Kruislaan 407, NL-1098SJ Amsterdam,  The Netherlands}
\address{$^2$Complex Photonic Systems, MESA$^+$ Institute for Nanotechnology,
University of Twente, The Netherlands}
\address{Corresponding author:  f.koenderink@amolf.nl}

\begin{abstract}
We have investigated the local density of optical states (LDOS) in
titania and silicon inverse opals -- three-dimensional photonic
crystals that have been realized experimentally. We used the H-field
plane-wave expansion method to calculate the density of states and
the projected local optical density of states, which are directly
relevant for spontaneous emission dynamics and strong coupling. We
present the first  quantitative analysis of the frequency resolution
and of the accuracy of the calculated local density of states. We
have calculated the projected LDOS for many different emitter
positions in inverse opals in order to supply a theoretical
interpretation for recent emission experiments and as reference
results for future experiments and theory by other workers. The
results show that the LDOS in inverse opals strongly depends on the
crystal lattice parameter as well as on the position and orientation
of emitting dipoles.
\end{abstract}

\ocis{160.5298,260.2510,270.5580,290.4210}

 ]  

\section{Introduction}

Photonic crystals are metamaterials with periodic variations of the
dielectric function on length scales comparable to the wavelength of
light. These dielectric composites are of a keen interest for
scientists and engineers because they offer exciting ways to
manipulate photons \cite{Joannopoulos95,Soukoulis01}. Of particular
interest are three-dimensional (3D) photonic crystals possessing a
photonic bandgap, \emph{i.e.\/} a frequency range where no  photon
modes exist at all. Photonic bandgap materials possess a great
potential for drastically  changing the rate of spontaneous emission
and for achieving localization of
light~\cite{Bykova,Bykovb,Yab87,John87}. Control over spontaneous
emission is important for many applications such as miniature
lasers~\cite{Painter99}, light-emitting diodes~\cite{Park04} and
solar cells~\cite{Graetzel01}. According to Fermi's golden rule, the
rate of spontaneous emission from quantum emitters such as atoms,
molecules or quantum dots is proportional to the `local radiative
density of states' (LDOS) \cite{Sprik96,Vats02}, which counts the
number of electromagnetic states at given frequency, location and
orientation of the dipolar emitters. In addition, nonclassical
effects  can occur that are beyond Fermi's Golden
Rule~\cite{Soukoulis01,Vats02}. These strong coupling phenomena,
such as fractional decay,   rely on coherent coupling of the quantum
emitter to a sharp feature in a highly structured LDOS. Whether
these effects  are in fact observable in real photonic crystals
depends on how rapid the LDOS changes within a small frequency
window~\cite{Kristensen08}.  Therefore  local density of states
calculations for experimentally realized photonic crystals are
essential to assess current spontaneous experiments and future
strong coupling experiments alike, provided that the calculations
are accurate and have a controlled frequency resolution.

LDOS effects on spontaneous emission in photonic crystals have been
experimentally demonstrated in a variety of systems. Since only
three-dimensional (3D) crystals promise full control over all
optical modes with which elementary emitters interact, many groups
have pursued 3D photonic crystals. Fabrication of such periodic
structures with high photonic strengths is, however, a great
challenge~\cite{Soukoulis01,Busch-Foell04}. Inverse opals are among
the most strongly photonic 3D crystals that can be fabricated
relatively easily using self-assembly methods. Such crystals consist
of \emph{fcc} lattices of close-packed air spheres in a backbone
material with a high dielectric
constant\cite{Wijnhovena,Wijnhovenb,Zakhidov98,Blanco00,Vlasov01}. In these
inverse opals, continuous-wave experiments on light sources with a
low quantum efficiency revealed inhibited radiative emission
rates~\cite{Koenderink02,Ogawa04}. Enhanced and inhibited
time-resolved emission rates have been observed from
highly-efficient emitters in 3D inverse opals~\cite{Lodahl04}.
Simultaneously, several groups have realized that emission
enhancement and partial inhibition   can also be obtained in 2D slab
structures \cite{Badolato05,Finley05,Noda05,Englund05,julsgaard08}.

Following the first calculations by Suzuki and Yu\cite{suzuki95},
several papers  have reported on calculations of the local density
of optical states in photonic crystals using both time
domain~\cite{lee00,hermann,Koenderink06} and frequency domain
methods~\cite{Busch98,Zhi-YLi00,Wang03,Fussell04}. Unfortunately,
analysis of experimental data for emitters in 3D crystals in terms
of these LDOS calculations is compounded by several problems in
literature:
\begin{enumerate}
\item Most prior calculations were performed for model systems that don't
correspond to structures used in experiments, and for emitter
positions that are not probed in experiments.
\item The accuracy of the reported LDOS has never been discussed, hampering comparison to experiments.
\item The frequency resolution of the reported LDOS has remained
unspecified. Therefore sharp features of relevance for non-classical
emission can not be assessed~\cite{Vats02,Kristensen08}.
\item  Many previously reported LDOS
calculations are erroneous for symmetry reasons, as pointed out by
Wang \emph{et al.} \cite{Wang03}.
\end{enumerate}

In this paper we aim to overcome all these problems. We benchmark
the accuracy and frequency resolution for our LDOS calculations to
allow quantitative comparison with experiments and with quantum
optics strong coupling requirements.  Since prior LDOS calculations are scarce
(and partly erroneous, item 1 above) we present sets of LDOS's for
experimentally relevant structures, and for spatial positions where
sources can be practically placed. Specifically, we model the
spatial distribution of the dielectric function
$\epsilon(\mathbf{r})$ in such a way that it closely resembles
$\epsilon(\mathbf{r})$ in titania (TiO$_2$)~\cite{Wijnhovena,Wijnhovenb} and
silicon (Si) inverse-opal photonic
crystals~\cite{Blanco00,Vlasov01}. For these two structures we
calculated the LDOS at various positions in the crystal unit cell
and for specific orientations of the transition dipoles. The results
on the TiO$_2$ inverse opals are relevant for interpreting recent
emission experiments \cite{Lodahl04,NikolaevPRB07}. To aid other
workers to interpret their experiments and to benchmark their codes,
we make data sets  that we report in
figures throughout this manuscript available as online material.

The paper is arranged as follows: in Section~\ref{sec:LDOScalc}, we
present a detailed description of the  method by which we have
calculated the photonic band structures and the LDOS. We discuss the
accuracy and frequency resolution of our calculations. In
Section~\ref{sec:comparison} we compare our computations with the
known DOS in vacuum and with earlier results on the DOS and
LDOS~\cite{Wang03} in 3D periodic structures. Section~\ref{sec:tio2} describes the
LDOS in inverse opals from TiO$_2$, and in Section~\ref{sec:Si} we
present results of the LDOS in Si inverse opal photonic band gap
crystals.

\section{Calculation of local density of states}\label{sec:LDOScalc}

\subsection{Introduction}\label{sec:LDOS-intro}

The local radiative density of optical states is defined as
\begin{equation}\label{LRDOS-PC}
N(\mathbf{r},\omega,\mathbf{e}_d)=
\frac{1}{(2\pi)^3}\sum_n\int_{BZ}d\mathbf{k}
\delta(\omega-\omega_{n,\mathbf{k}})|\mathbf{e}_{d}\cdot\mathbf{E}_{n,\mathbf{k}}(\mathbf{r})|^2,
\end{equation}
where integration over \textbf{k}-vector is performed over the first
Brillouin zone, \emph{n} is the band index and $\mathbf{e}_{d}$ is
the orientation of the emitting dipole.  The total density of states
(DOS) is the unit-cell and dipole-orientation average of the LDOS
defined as $N(\omega)= \sum_n\int_{BZ}d\mathbf{k}
\delta(\omega-\omega_{n,\mathbf{k}})$. The important quantities that
determine the LDOS are the eigenfrequencies $\omega_{n,\mathbf{k}}$
and electric field eigenmodes
$\mathbf{E}_{n,\mathbf{k}}(\mathbf{r})$ for each \textbf{k}-vector.
Calculation of these parameters will be discussed below in
Section~\ref{sec:PWmethod}.

The expression for the LDOS contains the term
$|\mathbf{e}_d\cdot\mathbf{E}_{n,\mathbf{k}}(\mathbf{r})|$ that
depends on the dipole orientation $\mathbf{e}_d$. It is important to
realize that in photonic crystals the vector fields
$\mathbf{E}_{n,\mathbf{k}}(\mathbf{r})$ are not invariant under the
lattice point-group operations $\alpha$, as first reported in
Ref.~\cite{Wang03}. Explicitly, this means that the projection of
the field of a mode at wave vector $\mathbf{k}$ on the dipole
orientation $\mathbf{e}_d$ (\emph{i.e.\/} %
$|\mathbf{e}_d\cdot\mathbf{E}_{n,\mathbf{k}}(\mathbf{r})|$)is not
identical to the projection
$|\mathbf{e}_d\cdot\mathbf{E}_{n,\alpha[\mathbf{k}]}(\mathbf{r})|$
of the symmetry related modes with wave vectors $\alpha[\mathbf{k}]$
on the same $\mathbf{e}_d$. As a consequence one can not calculate
the LDOS for a specific dipole orientation by restricting the
integral in Eq.~(\ref{LRDOS-PC}) over the irreducible part (1/48th)
of the Brillouin zone, since symmetry related wave vectors do not
give identical contributions. Unfortunately, in many previous
reports on the LDOS, this reduced symmetry for vector modes as
compared to scalar quantities was overlooked, resulting in erroneous
results~\cite{Wang03}.  In general, the only symmetry that can be
invoked to avoid using the full Brillouin zone for LDOS calculations
is inversion symmetry, which corresponds to time reversal symmetry.
Consequently, correct results require that exactly half of the
Brillouin zone is considered for LDOS calculations, rather than the
irreducible part of the Brillouin zone that was used in most
previous literature. We have explicitly verified  that our
implementation (using the $k_z\geq 0$ half of the Brillouin zone)
results in the same LDOS on symmetry related positions, provided
that one also takes the concomitant symmetry related dipole
orientation into account. Furthermore, our calculations confirm the
claim  by Wang et. al. that this required symmetry is only recovered
upon integration over half the Brillouin zone, rather than over just
the irreducible part as considered in, e.g.,
Ref.~\cite{Busch98,Zhi-YLi00}.

\subsection{Plane-wave expansion}\label{sec:PWmethod}

We use the H-field inverted plane wave expansion
method~\cite{Busch98,Ho90,Sozuer92} to solve for the electromagnetic
field modes in photonic crystals. For nonmagnetic materials, it is
most convenient to solve the wave equation for the
$\mathbf{H}(\mathbf{r})$ field~\cite{Jackson}
\begin{equation}\label{H-eigenvalue}
\nabla \times \left[\epsilon(\mathbf{r})^{-1}\nabla \times
 \mathbf{H}(\mathbf{r})\right] = \frac{\omega^2}{c^2}
 \mathbf{H}(\mathbf{r}).
 \end{equation}
 because the operator
$\nabla \times \epsilon(\mathbf{r})^{-1} \nabla \times$ is
Hermitian, and consequently has real eigenvalues ${\omega^2}/{c^2}$
\cite{Ho90,Sozuer92,Joannopoulos95,Busch98}. Because of the
periodicity of the dielectric function $\epsilon(\mathbf{r})$ in
photonic crystals, the field modes
$\mathbf{H}_{\mathbf{k}}(\mathbf{r})$ of the eigenvalue problem
Eq.~(\ref{H-eigenvalue}) satisfy the Bloch theorem \cite{Ashcroft}:
\begin{equation}
\mathbf{H}_{\mathbf{k}}(\mathbf{r})= e^{i\mathbf{k} \cdot
\mathbf{r}} \mathbf{u}_{\mathbf{k}}(\mathbf{r}).
\end{equation}
These Bloch modes are fully described by the wavevector \textbf{k}
and the periodic function $\mathbf{u}_{\mathbf{k}}(\mathbf{r})$,
which has the periodicity of the crystal lattice so that
$\mathbf{u}_{\mathbf{k}}(\mathbf{r})=\mathbf{u}_{\mathbf{k}}(\mathbf{r+R})$.
To solve the wave equation, the inverse dielectric function and the
Bloch modes are expanded in a Fourier series over the
reciprocal-lattice vectors \textbf{G}:
\begin{equation}\label{Fourier-exp1}
\epsilon(\mathbf{r})^{-1}=\eta(\mathbf{r})=\sum_\mathbf{G}
\eta_\mathbf{G} e^{i\mathbf{G}\cdot\mathbf{r}} \quad \mbox{and}
\end{equation}
\begin{equation}
\mathbf{H}_{\mathbf{k}}(\mathbf{r})=\sum_\mathbf{G}
\mathbf{u}^{\mathbf{k}}_\mathbf{G} e^{i(\mathbf{k}+\mathbf{G})\cdot
\mathbf{r}},
\end{equation}
where $\eta_\mathbf{G}$ and $\mathbf{u}^{\mathbf{k}}_\mathbf{G}$ are
the 3D Fourier expansion coefficients of respectively
$\eta(\mathbf{r})$ and $\mathbf{u}_{\mathbf{k}}(\mathbf{r})$.
Substituting these expressions into the H-field wave equation in
Eq.~(\ref{H-eigenvalue}), we obtain a linear set of eigenvalue
equations:
\begin{equation}\label{planewave2}
-\sum_{\mathbf{G'}} \eta_\mathbf{G-G'} (\mathbf{k+G})\times
[(\mathbf{k+G'}) \times \mathbf{u}^{n,\mathbf{k}}_\mathbf{G'}]
=\frac{{\omega_n^2(\mathbf{k})}}{c^2}
\mathbf{u}^{n,\mathbf{k}}_\mathbf{G}.
\end{equation}%
This infinite equation set with the known parameters \textbf{G} and
$\eta_\mathbf{G-G'}$ determines all allowed frequencies
$\omega_n(\mathbf{k})$ for each value of the wave vector \textbf{k},
subject to the transversality requirement $\nabla \cdot
\mathbf{H}_{\mathbf{k}}(\mathbf{r})=0$. Due to the periodicity of
$\mathbf{u}_{\mathbf{k}}(\mathbf{r})$, we can restrict \textbf{k} to
the first Brillouin zone. For each wave vector \textbf{k}, there is
a countably infinite number of modes with discretely spaced
frequencies. All the modes are labeled with the band number \emph{n}
in order of increasing frequency and are described as a family of
continuous functions $\omega_n(\mathbf{k}$) of $\mathbf{k}$.

To compute the eigenfrequencies $\omega_n(\mathbf{k})$ and the
expansion coefficients of the eigenmodes
$\mathbf{u}^{n,\mathbf{k}}_\mathbf{G}$, the infinite equation set is
truncated. By restricting the number of reciprocal-lattice vectors
\textbf{G} to a finite set $\mathcal{G}$ with $N_G$ elements,
Eq.~(\ref{planewave2}) is limited to a $3N_G$ dimensional equation
set. In our implementation we choose the truncated set $\mathcal{G}$
to correspond to the set of all reciprocal lattice vectors within a
sphere centered around the origin of k-space. The transversality of
the H-field gives an additional condition on the eigenmodes:
$(\mathbf{k+G})\cdot \mathbf{u}^{n,\mathbf{k}}_\mathbf{G}$ = 0,
which eliminates one vector component of
$\mathbf{u}^{n,\mathbf{k}}_\mathbf{G}$. Following
Ref.~\cite{Busch98}, for each $\mathbf{k+G}$ one needs to find two
orthogonal unit vectors $\mathbf{e}^{1,2}_\mathbf{k+G}$ that form an
orthogonal triad with $\mathbf{k+G}$. By expressing the eigenmode
expansion coefficients in the plane normal to $\mathbf{k+G}$ as
$\mathbf{u}^{n,\mathbf{k}}_\mathbf{G}=u^{n,\mathbf{k}}_{\mathbf{G},1}\mathbf{e}^{1}_\mathbf{k+G}
+u^{n,\mathbf{k}}_{\mathbf{G},2}\mathbf{e}^{2}_\mathbf{k+G}$, we
remove one third of the unknowns. Then, Eq.~(\ref{planewave2})
becomes
\begin{eqnarray}\lefteqn{\sum_{\mathbf{G'}\in
\mathcal{G}}\eta_\mathbf{G-G'}
|\mathbf{k+G}||\mathbf{k+G'}|\cdot\Bigg[} \nonumber \\ &&
 \left(\begin{array}{c c}
\mathbf{e}^{2}_\mathbf{k+G}\cdot \mathbf{e}^{2}_\mathbf{k+G'} & -\mathbf{e}^{2}_\mathbf{k+G}\cdot \mathbf{e}^{1}_\mathbf{k+G'} \\
-\mathbf{e}^{1}_\mathbf{k+G}\cdot \mathbf{e}^{2}_\mathbf{k+G'} &
\mathbf{e}^{1}_\mathbf{k+G}\cdot \mathbf{e}^{1}_\mathbf{k+G'}
\end{array}\right)\left(\begin{array}{c}
u^{n,\mathbf{k}}_{\mathbf{G'},1} \\
u^{n,\mathbf{k}}_{\mathbf{G'},2}
\end{array}\right)\Bigg]
 \nonumber \\ &&
 \qquad \qquad \qquad \qquad =\frac{{\omega_n^2(\mathbf{k})}}{c^2} \left(\begin{array}{c}
u^{n,\mathbf{k}}_{\mathbf{G},1} \\
u^{n,\mathbf{k}}_{\mathbf{G},2}
\end{array}\right), \quad \forall~ \mathbf{G} \in
\mathcal{G}.\nonumber \\  \label{planewave3}
\end{eqnarray}
To find the matrix of Fourier coefficients $\eta_\mathbf{G-G'}$, we
used the method of Refs.~\cite{Ho90,Busch98}. The coefficients are
computed by first Fourier-transforming the dielectric function
$\epsilon(\mathbf{r})$, and then truncating and inverting the
resulting matrix. As first noted by Ho, Chan and
Soukoulis\cite{Ho90}, using the inverse of $\epsilon_\mathbf{G-G'}$
rather than $\eta_{\mathbf{G-G'}}$ dramatically improves the (poor)
convergence of the plane-wave method that is associated with the
discontinuous nature of the dielectric function~\cite{Sozuer92}. A
rigorous explanation for this improvement was put forward by
Li~\cite{Li96}, who studied the presence of Gibbs oscillations in
the truncated Fourier expansion of products of functions with
complementary jump discontinuities. Using the H-field inverted
matrix plane wave method, the frequencies obtained with N$_G$ = 725
(for \emph{fcc} structures) deviate by less than 0.5~\% from the
converged band structures~\cite{Busch98}. Solving
Eqs.~(\ref{planewave3}) gives the frequencies $\omega_n(\mathbf{k})$
and the Fourier expansion coefficients for the H-field eigenmodes
$\mathbf{H}_{n,\mathbf{k}}\mathbf{(r)}$ needed to calculate the LDOS
in the photonic crystal. The required E-fields
$\mathbf{E}_{n,\mathbf{k}}\mathbf{(r)}$ are obtained using the
Maxwell equation  $\partial \mathbf{D}/\partial t = \nabla\times
\mathbf{H}$:
\begin{eqnarray}
\mathbf{E}_{n,\mathbf{k}}\mathbf{(r)}& =&
\frac{1}{\omega_n(\mathbf{k})\epsilon_0}\sum_{\mathbf{G,G'}\in\mathcal{G}}
\eta_\mathbf{G'-G}|\mathbf{k+G}|\cdot \bigg[\qquad \nonumber \\
& &\qquad
\left(u^{n,\mathbf{k}}_{\mathbf{G},1}\mathbf{e}^{2}_\mathbf{k+G}
-u^{n,\mathbf{k}}_{\mathbf{G},2}\mathbf{e}^{1}_\mathbf{k+G}\right)e^{i(\mathbf{k+G'})\cdot
\mathbf{r}}\bigg]. \nonumber \\ \label{eq:ecalcfromh}
\end{eqnarray}
From the orthonormality of the eigenvectors of
Eq.~(\ref{planewave3}) it follows that the Bloch functions
$\mathbf{H}_{n,\mathbf{k}}\mathbf{(r)}$ and
$\mathbf{E}_{n,\mathbf{k}}\mathbf{(r)}$ defined above satisfy the
orthonormality relations:
\begin{equation}
\int_{BZ}\mathbf{H}_{n,\mathbf{k}}\mathbf{(r)}\cdot\mathbf{H}^*_{n',\mathbf{k'}}\mathbf{(r)}d\mathbf{r}=\delta(\mathbf{k}-\mathbf{k'})\delta_{n,n'},
\end{equation}
\begin{equation}
\int_{{BZ}}\epsilon\mathbf{(r)}\mathbf{E}_{n,\mathbf{k}}\mathbf{(r)}\cdot\mathbf{E}^*_{n',\mathbf{k'}}\mathbf{(r)}d\mathbf{r}=\delta(\mathbf{k}-\mathbf{k'})\delta_{n,n'}.
\end{equation}
It should be noted in the definition of $\mathbf{E}_{n,\mathbf{k}}$
that the multiplication by $1/\epsilon(\mathbf{r})$ to calculate
$\mathbf{E}$ from $\mathbf{D}$ is not in the denominator in front of
the Fourier expansion. Rather it appears as a matrix multiplying the
$\mathbf{D}$-field in Fourier space, i.e., within the sum over
reciprocal lattice vectors. This ordering ensures that complementary
jumps in $\mathbf{D}$ and $1/\epsilon(\mathbf{r})$ cancel, even for
the truncated Fourier series, as can be easily checked by plotting
calculated mode profiles for TM modes in 2D crystals.

\subsection{Frequency resolution and accuracy of  LDOS}\label{ssec:accur_&_resol}
We are not aware of any previous report that benchmarks the accuracy
of the calculated local radiative density of states or that
specifies the frequency resolution. Motivated by the requirements
for accurate results for comparison with experiments and for judging
the utility of crystals for non-classical emission dynamics, we
consider the accuracy and resolution of our approximation to the
LDOS integral in Eq.~(\ref{LRDOS-PC}). The main approximation is to
replace the integration over wave vector \textbf{k}  by an
appropriately weighted summation over a discrete set of wave vectors
on a discretization grid\cite{MonkhorstPack}. Either interpolation
schemes~\cite{Busch98} or simple histogramming (`root-sampling')
methods are used to compute the LDOS. The k-point grid density sets
the number of k-points in the Brillouin zone, $N_k$. The accuracy of
the resulting LDOS approximation is set by the the density of grid
points that is used to discretize the wave vector integral. Due to
the transparent relation between the accuracy of the DOS, the
frequency resolution and the \textbf{k}-vector sampling resolution,
we will focus on the simple histogramming approach. For the LDOS
computations  one chooses a certain frequency bin width
$\Delta\omega$ to build an LDOS histogram. For a desired frequency
resolution $\Delta \omega$ and a desired accuracy for the LDOS
content $N(\mathbf{r},\omega,\mathbf{e}_d)\Delta \omega$ in each
frequency bin, one needs to choose an appropriate \textbf{k}-vector
spacing $\Delta k$ that depends on the steepness of the sampled
dispersion relation $\omega(\mathbf{k})$.
\begin{figure}
\centerline{\includegraphics[width=\figwidth]{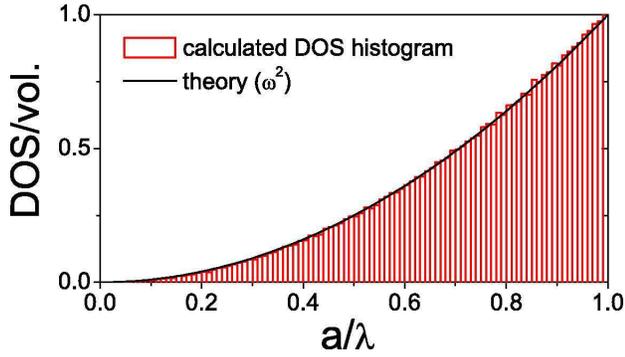}}
\caption{\label{vacuumLDOS} (Color online) DOS per volume   in units
$4/a^2c$ for vacuum modeled as an empty \emph{fcc} crystal. The
calculated DOS shown by red histogram bars is compared to the
analytically derived $\omega^2$ behavior (curve). In vacuum the DOS
per volume equals the dipole-averaged LDOS. }
\end{figure}
\begin{figure}
\centerline{\includegraphics[width=\figwidth]{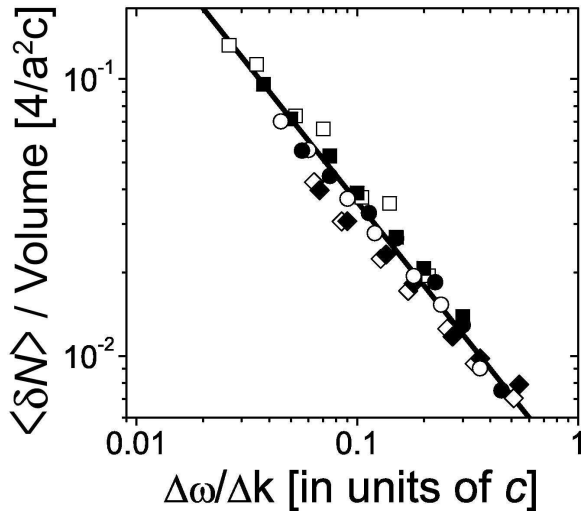}}
\caption{\label{DOSerror} Average absolute deviation of the
calculated DOS from the exact total DOS $N(\omega)$ for an `empty
crystal.' The average runs over the frequency range $0<\omega<2\pi
c/a$ and the deviation is in units of the DOS $N(\omega)$ per volume
at $\omega\, a/2\pi c = 1$, i.e., in units $4/a^2 c$. In accordance
with Eq.~(\ref{histbin}), the error is inversely proportional to the
ratio $\Delta \omega/\Delta k$ of the histogram bin width $\Delta
\omega$ to the integration grid spacing $\Delta k$. Symbols
correspond to integration using the number of k-points $N_k$ = 280,
770, 1300, 2480, 2992 and 3570
($\square,\blacksquare,\circ,\bullet,\lozenge,\blacklozenge$)
respectively, with various $\Delta \omega$.}
\end{figure}
Indeed, the useful frequency resolution $\Delta\omega$ of a
histogram of the DOS and LDOS is limited by the resolution $\Delta
k$ of the grid in $\mathbf{k}$ space to be
\begin{equation}\label{histbin}
\Delta \omega \approx \Delta k |\nabla_{\mathbf{k}} \omega|,
\end{equation}
as detailed in Ref.~\cite{gilat72} for the electronic DOS. This
criterion relates the separation between adjacent $\mathbf{k}$-grid
points to their approximate frequency spacing via the group
velocity. If histogram bins are chosen too narrow  compared to the
expected frequency spacing between contributions to the discretized LDOS
integral, unphysical spikes appear in the approximation, especially
in the limit of small $\omega$, where the group velocity\index{group
velocity} $|\nabla_{\mathbf{k}} \omega |$ is usually largest. Apart
from full gaps in the LDOS, photonic crystals also promise sharp
lines at which the LDOS is enhanced, which are important for
non-classical emission dynamics. Hence it is especially important to
distinguish sharp spikes that are due to histogram binning noise
from true features. Unfortunately, many reports in literature
feature sharp spikes that are evidently binning noise (wave vector
undersampling errors), as they occur in the long wavelength limit,
below any stop gap.

To improve the resolution without adding time-consuming
diagonalizations, several interpolation schemes have been
suggested~\cite{gilat72}. Within the histogramming approach, an
interpolation scheme essentially improves binning statistics by
adding histogram contributions from intermediate grid points on the
assumption that quantities vary linearly between grid points. While
interpolation decouples the binning noise from the frequency bin
width $\Delta \omega$, resulting in arbitrarily smooth LDOS
approximations, it has the disadvantage of obscuring the intrinsic
relation between frequency resolution and wave vector resolution in
Eq.~(\ref{histbin}).

A good benchmark for the binning noise of the $\mathbf{k}$-space
integration, independent of the convergence of the plane-wave
method, is to calculate the LDOS or DOS of an `empty' crystal, with
uniform dielectric constant equal to unity. Such an `empty' crystal
represents a limit of zero photonic strength and the maximum
possible group velocity $|\nabla_\mathbf{k} \omega|$. In
Figure~\ref{vacuumLDOS} we show the DOS in an empty \emph{fcc}
crystal. As expected for a crystal with zero dielectric contrast,
the calculated DOS is independent of dipole position and
orientation. In agreement with the well-known DOS in vacuum the
calculated  DOS increases parabolically with frequency. Fluctuations
around the parabola are due to binning noise associated with the
finite k-space discretization.

We have calculated the relative root-mean-square error in the
calculated density of states (DOS) for an \emph{fcc} `empty' crystal
averaged over all histogram bins in the frequency range $0<\omega a
/2\pi c <1$ for several combinations of histogram binwidth and
k-grid resolution, as specified in the caption of
Fig.~\ref{DOSerror}. As predicted by Eq.~(\ref{histbin}), the
deviation of the approximation from the analytic result is inversely
proportional to the ratio $\Delta \omega / \Delta k$. In most cases,
one wants to calculate the LDOS with a given frequency resolution
$\Delta \omega$, \emph{i.e.}
$N(\mathbf{r},\omega,\mathbf{e}_d)\Delta \omega$, to within a
predetermined accuracy. For instance, calculating the vacuum DOS for
frequencies $0<\omega a /2\pi c <1$  with a desired absolute
accuracy better than $0.01\cdot 4/a^2 c$ (1\% of the vacuum DOS at
$\omega a /2\pi c =1$) and a desired frequency resolution of
$\Delta\omega=0.01 (2\pi c/a)$, requires using $\Delta k \sim \Delta
\omega/0.3c$. The number of k-points corresponding to this wave
vector sampling equals 2480 k-points in the irreducible wedge of
the \emph{fcc} Brillouin zone, 59520 in the required half Brillouin
zone, or equivalently $N_k=119040$ in the full Brillouin zone.
Photonic crystals with nonzero index contrast cause a pronounced
frequency structure of the LDOS. Due to the flattening of bands
compared to the dispersion bands of vacuum-only crystals, the
k-space integration itself is  at least as accurate, assuming that
the eigenfrequencies and the field-mode patterns are known with
infinite accuracy. In practice, one needs to adjust the number of
plane waves to obtain all eigenfrequencies to within the desired
frequency resolution $\Delta\omega$. In our computations for
\emph{fcc} crystals, we represented the k-space of a half of the
first Brillouin zone by an equidistant grid consisting of
$N_k/2=145708$ k-points. The frequency resolution of our LDOS
histograms is $\Delta \omega = 0.01\cdot 2\pi c/a$, and the spatial
resolution of the LDOS is approximately $a/40$ judging from the
maximum $|\mathbf{G}-\mathbf{G'}|\approx 120/a$ involved in the
expansion with $N_G$ = 725 plane waves in Eq.~(\ref{Fourier-exp1}).

\subsection{Computation time required for LDOS} 
Calculating the LDOS in 3D periodic structures is a time-consuming
task. The chosen degree of k-space discretization ($N_k/2 = 145708$
k-points) and the number of plane-waves ($N_G$ = 725) are the result
of a trade-off between the desired accuracy and tolerated duration
of the calculations. Essentially the computation consists of solving
for the lowest $n$ eigenvalues and eigenvectors of a real symmetric
$2N_G \times 2N_G$ matrix for each of the $N_k$ k-points
independently. To accomplish this calculation, we use the standard
Matlab "eigs" implementation \cite{matlab} of an implicitly
restarted Arnoldi method (ARPACK) that takes approximately 2.2
seconds to find the lowest 20 eigenvalues and eigenvectors of a
single Hermitian 1500x1500 (i.e., $N_G\approx 750$ plane waves)
matrix on a 3 GHz Intel Pentium 4 processor, or on a 2.4 GHz Intel
Core Duo processor. For $145708$ k-points the resulting
computation time is hence on the order of 90 hours (4 days). Since
the Matlab ARPACK routine is already highly optimized, we do not
expect that the computation time per k-point can be significantly
reduced for LDOS calculations based on the standard H-field inverted
matrix plane wave method. In terms of the number of plane waves
$N_G$, the computation time scales as $N_G^3$.  As in
Ref.~\cite{sgjohnson01}, the algorithm can be accelerated by
realizing that the iterative eigensolver does not require the full
matrix ${\cal H}$ multiplying the vector $\mathbf{u}$ in
Eq.~\ref{planewave3}, but rather a function that quickly computes
the image ${\cal H} \mathbf{u}$ of any trial vector $\mathbf{u}$.
Since calculating ${\cal H} \mathbf{u}$ repeatedly involves a slow
matrix-vector multiplication with $\eta_{\mathbf{G}-\mathbf{G'}}$, the algorithm is
only accelerated for $N_G > 1000$.

\section{Comparison with previous results}\label{sec:comparison}

\begin{figure}
\centerline{\includegraphics[width=\figwidth]{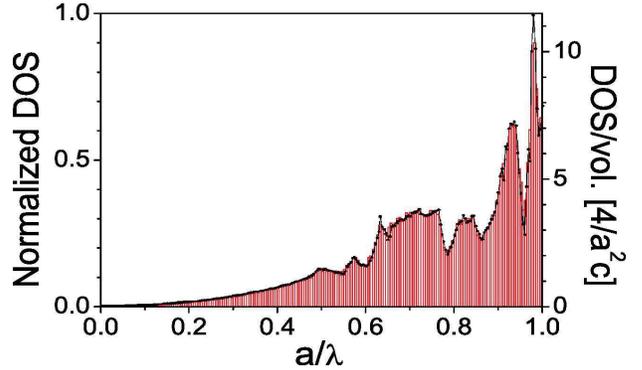}}
  \caption{\label{Busch98-DOS} (Color online)DOS per volume
in an \emph{fcc} crystal consisting of spheres with $\epsilon$ =
7.35 in a medium with $\epsilon$~=~1.77 with a filling fraction of
the spheres of 25 vol~\%. The solid dotted curve represents
calculations from Ref.~\cite{Busch98}. Our result is plotted as a
histogram (red).}
\end{figure}

\begin{figure}
\centerline{\includegraphics[width=\figwidth]{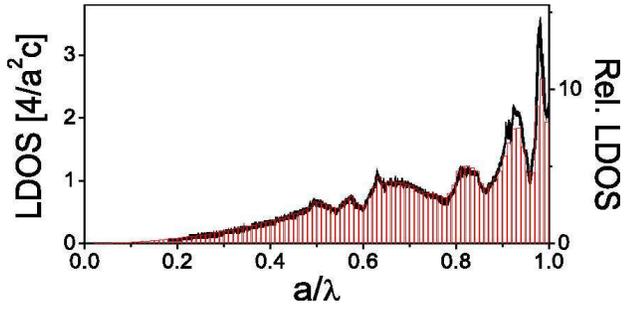}}
  \caption{\label{Wang03-LDOS} (Color online)
Dipole-averaged LDOS in the same photonic crystal as in
Fig.~\ref{Busch98-DOS} at a position $(\frac{1}{4}, \frac{1}{4},
0)$. Red histogram: our calculations (Media 1). Solid curve: results
from Ref.~\cite{Wang03}. This relative LDOS is the ratio of the LDOS
to that in vacuum at $a/\lambda$ = 0.495.}
\end{figure}

To test the computations, we compare our results with earlier
reports. We have calculated the DOS and LDOS in an \emph{fcc}
crystal consisting of dielectric spheres with $\epsilon_1$ = 7.35
(TiO$_2$) in a medium with $\epsilon_2$ = 1.77 (water) -- the same
structure  as was analyzed in Refs~\cite{Busch98,Wang03}. The
spheres occupy 25 vol\% of the crystal. Figure~\ref{Busch98-DOS}
shows the total DOS in this photonic crystal calculated by us  and
by Busch and John~\cite{Busch98} . We reproduce the earlier
calculations of the total DOS: both results shown in
Fig.~\ref{Busch98-DOS} are in excellent agreement, with deviations
less than 2\% throughout the frequency range $0<\omega a/2\pi c<1$.
In Figure~\ref{Wang03-LDOS} (Media 1)we demonstrate the LDOS in the
same photonic crystal at a specific location: at a point equidistant
from two nearest-neighbor spheres. In this calculation, we used the
same number of reciprocal-lattice vectors N$_G$ = 965 as in the only
available benchmark paper by Wang et al.~\cite{Wang03} that does not
contain symmetry errors. We find that our calculations (empty
circles) are in good agreement up to $a/\lambda=0.85$ with the LDOS
reported previously (solid curve): the deviations are smaller than
1\%. Deviations at higher frequencies are either due to a
difference in accuracy of k-space integration (frequency binning
resolution and k-space sampling density), or to a difference in
accuracy of the plane wave methods (eigenmodes and
eigenfrequencies).
Unfortunately, neither the accuracy nor the
frequency resolution is specified in Ref.~\cite{Wang03}. The k-space
sampling density in our work is approximately twice the density
specified in Ref.~\cite{Wang03}. Based on the excellent agreement of
our total DOS calculations with those of Busch and
John~\cite{Busch98}, and on the fact that Wang et al. used a k-space
density and plane wave number comparable to that of Busch and John,
it is unlikely that inaccuracy of the k-space integration in either
calculation is the source of discrepancy. We therefore surmise that
deviations are due to a difference in the method of evaluation of
$\mathbf{E}_{n,\mathbf{k}}(\mathbf{r})$. Conversion of $\mathbf{D}$
to  $\mathbf{E}$ by multiplication with $1/\epsilon(\mathbf{r})$ in
real space as proposed in~\cite{Wang03} can cause incorrect mode
amplitudes due to Gibbs oscillations, as opposed to the Fourier
space conversion using Eq.~(\ref{eq:ecalcfromh}). In general our
calculations confirm the result by Wang et. al.~\cite{Wang03} that
the LDOS is only correctly calculated by integration over half the
Brillouin zone, rather than over the irreducible part as was
incorrectly used in earlier reports.

\section{LDOS in TiO$_2$ inverse opals}\label{sec:tio2}

\begin{figure}
\includegraphics[width=\figwidth]{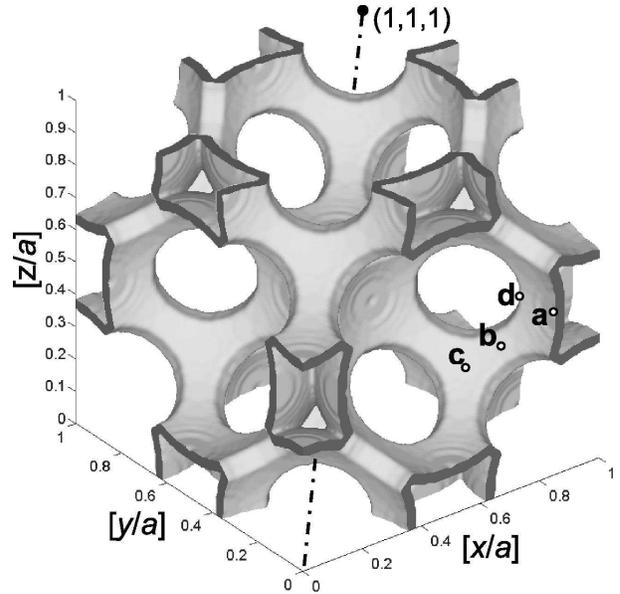}
  \caption{\label{inverse-opal-unit-cell} Rendering of the
dielectric function in one \emph{fcc} unit cell that models the
TiO$_2$ inverse-opal structure in section~\ref{sec:tio2}: an
\emph{fcc} lattice of air spheres of radius $r=0.25\sqrt{2}a$ with
$a$ being the cubic lattice parameter. The spheres are covered by
shells with $\epsilon = 6.5$ and outer radius 1.09$r$. Neighboring
air spheres are connected by windows of radius 0.4\emph{r}. The
letters (\textbf{a}--\textbf{d}) indicate four different positions
at the TiO$_2$-air: \textbf{a} = $(1,0,0)/(2\sqrt{2})$, \textbf{b} =
$(1,1,2)/(4\sqrt{3})$, \textbf{c} = $(1,1,1)/(2\sqrt{6})$ and
\textbf{d} = (0.33,0.13,0) (points shown are symmetry-equivalents).
The dash-dotted line shows the main diagonal of the cubic unit
cell.}
\end{figure}
\begin{figure}
\includegraphics[width=\figwidth]{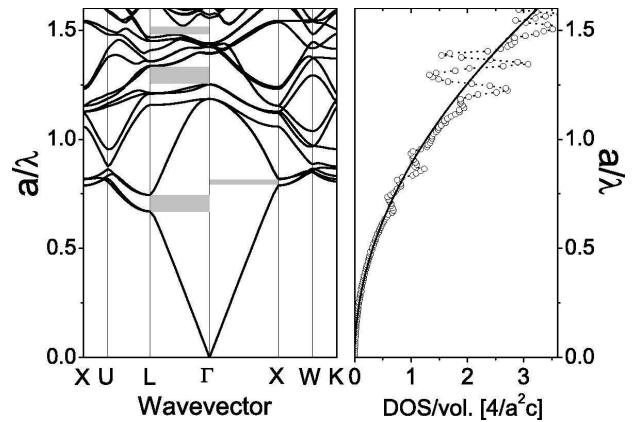}
 \caption{{\emph{Left}}: Photonic band structure  (Media
2)for the TiO$_2$ inverse opal shown in
Fig.~\ref{inverse-opal-unit-cell}. The grey rectangles indicate
stopgaps in the $\Gamma$L direction and one stopgap in the $\Gamma$X
direction for the inverse opal. The stopgaps result in the decreased
DOS (\emph{right}: circles, Media 3) at corresponding frequencies
compared to the DOS in a homogeneous medium with $n_{av}= 1.27$
(\emph{right}: solid line).\label{bandstruct}}
\end{figure}

In recent time-resolved experiments, enhanced and inhibited emission
rates were demonstrated for quantum dots embedded inside strongly
photonic TiO$_2$ inverse
opals~\cite{Koenderink02,Lodahl04,NikolaevPRB07}. In the framework
of these experiments, it is highly relevant to calculate the LDOS
inside such inverse-opal photonic crystals, especially for the
source positions occupied in experiments. We model the position
dependence of the dielectric function $\epsilon(\mathbf{r})$ as
shown in Figure~\ref{inverse-opal-unit-cell}. This model assumes an
infinite \emph{fcc} lattice of air spheres with radius
$r=0.25\sqrt{2}a$ (\emph{a} is the cubic lattice parameter). The
spheres are covered by overlapping dielectric shells ($\epsilon =
6.5$) with outer radius 1.09\emph{r}. Neighboring air spheres are
connected by cylindrical windows of radius 0.4\emph{r}. The
resulting volume fraction of TiO$_2$ is equal to about 10.7\%. These
structural parameters are inferred from detailed characterization of
the inverse opals using electron microscopy and small angle X-ray
scattering~\cite{Wijnhovena,Wijnhovenb}. Moreover, the stop gaps in
the photonic band structure (Fig.~\ref{bandstruct} (Media 2 and
Media 3)) calculated using this model agree well with reflectivity
measurements in the ranges of both the first-order ($a/\lambda=0.7$)
and second-order Bragg diffraction ($a/\lambda=1.2$) \cite{Vos00}.

\begin{figure}
\includegraphics[width=\figwidth]{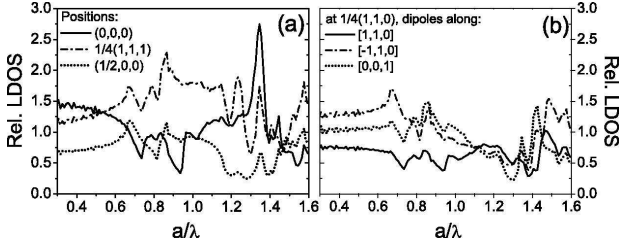}
  \caption{\label{LDOS-in-TiO2-1} Relative LDOS in the
inverse opal shown in Fig~\ref{inverse-opal-unit-cell} at three
different positions: (a, Media 4) $\mathbf{r}$ = (0,0,0) [the center
of an air sphere, solid curve], $\mathbf{r}$ = $\frac{1}{4}(1,1,1)$
[among three air spheres, dash-dotted curve] and $\mathbf{r}$ =
($\frac{1}{2}$,0,0) [midway between two spheres along [1,0,0]
direction, dotted curve]; (b, Media 5) $\mathbf{r}$ =
$\frac{1}{4}(1,1,0)$ [in the window between two spheres] projected
on [1,1,0], [-1,1,0] and [0,0,1] directions shown by solid,
dash-dotted and dotted curves, respectively.}
\end{figure}

Henceforth, we will consider the relative LDOS, which is the ratio
of the LDOS in a photonic crystal to that in a homogeneous medium
with the same volume-averaged dielectric function. This scaling is
motivated by experimental practice, in which emission rate
modifications are judged by normalizing measured rates to  the
emission rate of the same emitter in crystals with much smaller
lattice constant $a$~\cite{Koenderink02,Lodahl04,NikolaevPRB07}. In
these reference systems, emission frequencies correspond to  the
effective medium limit $a/\lambda \ll 0.5$ quantified by an average
index $n_{av}=\sqrt{\epsilon_{av}}= 1.27$ for the TiO$_2$
crystals~\cite{Koenderink02}. In units of $4/a^2c$, the LDOS in a
homogeneous medium is equal to $n_{av}(a/\lambda)^2/3$. In
Figure~\ref{LDOS-in-TiO2-1}a (Media 4) we plot the resulting LDOS at
three positions in the unit cell: at \textbf{r} = (0,0,0),
$\frac{1}{4}(1,1,1)$ and ($\frac{1}{2}$,0,0). Due to the high
symmetry of these points, the LDOS does not depend on the dipole
orientation, as we verified explicitly. A first main observation
from Fig.~\ref{LDOS-in-TiO2-1}a (Media 4) is that the LDOS differs
considerably between these three positions at all reduced
frequencies. This observation illustrates the well-known strong
dependence of the LDOS on position within the unit cell of photonic
crystals~\cite{suzuki95,Sprik96,Busch98}.

A second main observation in Fig.~\ref{LDOS-in-TiO2-1}a (Media 4) is
that the LDOS strongly varies with reduced frequency, revealing
troughs and peaks caused by the pseudogap near $a/\lambda = 0.7$,
which is related to $1^{st}$-order stop gaps such as the L-gap. The
effects of $2^{nd}$-order stopgaps appear beyond $a/\lambda >
1.15$~\cite{Vos00}. In the middle of the air region at position
\textbf{r} = (0,0,0) there is a sharp, factor-of-three enhancement
at $a/\lambda \approx 1.35$ within a  narrow frequency range. This
feature could be probed by resonant atoms infiltrated in the
crystals~\cite{Harding08}. At position \textbf{r} =
($\frac{1}{2}$,0,0) in an interstitial, the mode density has a broad
trough near $a/\lambda = 1.25$ that will lead to strongly inhibited
emission.

A third main observation is that at spatial positions with low
symmetry, the  LDOS clearly depends on the orientation of the
transition dipole moment. Figure~\ref{LDOS-in-TiO2-1}b (Media 5)
shows the frequency dependent LDOS for three perpendicular dipole
orientations at a position in the center of a window that connects
two neighboring air spheres (see
Figure~\ref{inverse-opal-unit-cell}). The LDOS differs for all three
orientations, and is thus anisotropic. At low frequency ($a/\lambda
= 0.3$), the emission rate is highest for a dipole pointing in the
$(1,1,0)$ direction, with increasing frequency the highest rate
shifts to the $(0,0,1)$ and then to the $(-1,1,0)$ orientation. We
emphasize that for emitters with fixed or slowly varying dipole
orientations, such as dye molecules or quantum dots on solid
interfaces, the emission rate is determined by the optical modes
that are projected on the dipole orientation. Therefore, knowledge
of the projected LDOS is important for controlling
spontaneous-emission rates as well as for interpreting the data from
experiments on emitters in photonic metamaterials.

A fourth main observation from Fig.~\ref{LDOS-in-TiO2-1}a and b
(Media 4 and 5)  is that at low frequencies $a/\lambda < 0.5$, the
relative LDOS hardly varies with frequency, which means that the
mode density is proportional to $\omega^2$, as in homogeneous media.
Interestingly, there is a clear dependence of the LDOS on both the
position and the dipole orientation even at these low frequencies,
\emph{i.e.}, long wavelengths relative to the crystal periodicity.
The reason for these effects is that the photonic Bloch modes
exhibit local variations of the electric field related to local
variations of the dielectric function in order to satisfy the
continuity equations at dielectric boundaries for the parallel
$\mathbf{E}$ or perpendicular $\mathbf{D}$ field,
respectively~\cite{Fussell04,Rogobete}. Consequently, the LDOS
strongly varies on length scales much less than the wavelength,
i.e., even in electrostatic or effective medium limit. While such
behavior may appear surprising, its origin in electrostatic
depolarization effects has been discussed
before~\cite{Miyazaki98,Rogobete}.

\begin{figure}
\includegraphics[width=\figwidth]{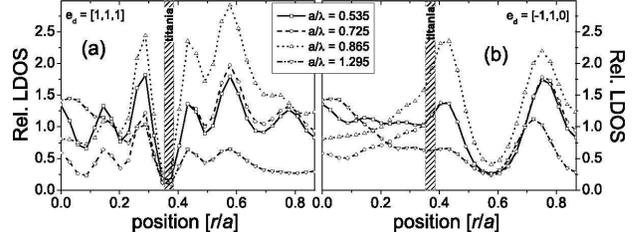}
\caption{\label{LDOS-vs-position-along111} Relative LDOS at four key frequencies, $a/\lambda$ = 0.535, 0.725, 0.865 and
1.295, in the inverse opal as a function of position \emph{r}
from (0,0,0) in the [1,1,1] direction. The hatched boxes indicate the position of the dielectric TiO$_2$ shell. The LDOS is projected on two dipole orientations: (a) [1,1,1] perpendicular to the dielectric-air interface, and (b) [-1,1,0] parallel to the interface. The LDOS projected on the [-1,-1,2] and [-1,1,0] directions is equal. For $r/a \in
[\frac{\sqrt{3}}{2},\sqrt{3}]$ the LDOS is mirror-symmetric to that
in the region [$0,\frac{\sqrt{3}}{2}]$.}
\end{figure}
To gain more insight in the spatial dependence of the LDOS in the
inverse opals,  we have performed calculations for dipoles
positioned along a characteristic axis in the unit cell.
Figure~\ref{LDOS-vs-position-along111} shows the LDOS at four key
frequencies for dipoles placed on the body diagonal of the cubic
unit cell, that is, from $(0,0,0)$ in the $[1,1,1]$ direction. The
LDOS has clear maxima and minima along the diagonal, varying by more
than $10 \times$. Figure~\ref{LDOS-vs-position-along111}(a) shows
that for a dipole oriented in the $[1,1,1]$ direction (perpendicular
to the dielectric-air interface), the LDOS is strongly ($>$~5x)
suppressed near the dielectric TiO$_2$ shell ($\emph{r/a} \approx
0.353$) at all frequencies. In contrast, no strong suppression or
enhancement occurs for dipole orientations parallel to the
dielectric interface, see Fig.~\ref{LDOS-vs-position-along111}(b).
The strongly differing LDOS for dipoles near the dielectric
rationalize the broad distributions of emission rates that were
recently observed for quantum dots in inverse
opals~\cite{NikolaevPRB07}, see below. Enhancements and inhibitions
also occur in the air regions: the LDOS is enhanced at $\emph{r/a}
\approx$ 0.28 and 0.57 by up to 2.5 to 3 times, respectively, see
Figure~\ref{LDOS-vs-position-along111}(a). At point $\emph{r/a} =
1/\sqrt{3} \approx 0.57$ that lies in the (111) lattice plane in the
air region, the LDOS is inhibited at all frequencies for the dipole
orientations [-1,1,0] and [-1,-1,2] that are perpendicular to the
(111) plane (see Fig.~\ref{LDOS-vs-position-along111}(b)). Finally,
for dipoles parallel to the body diagonal
(Figure~\ref{LDOS-vs-position-along111}(a)) the mode density shows
pseudo-oscillatory behavior on length scales much less than the
wavelength, \emph{e.g.}, a period of $0.2a$ at frequency $0.535
a/\lambda$ (period corresponds to $1/10^{th}$ of a wavelength). This
observation confirms that photonic crystals are \emph{bona fide}
metamaterials where optical properties strongly vary on length
scales much less than the wavelength.

\begin{figure}
\includegraphics[width=\figwidth]{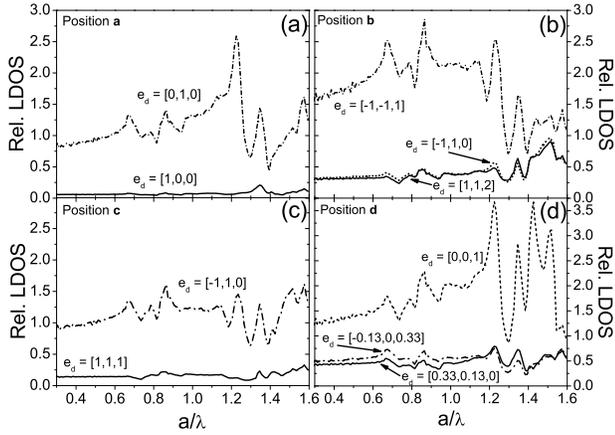}
\caption{\label{LDOS-in-TiO2-on-shell} Relative LDOS in
the inverse opal at four different positions on the TiO$_2$-air
interface shown in Fig.~\ref{inverse-opal-unit-cell}. At each
position the LDOS is projected on three mutually orthogonal dipole
orientations $\mathbf{e}_d$. (a, Media 6) point \textbf{a} for
$\mathbf{e}_d$ = [1,0,0] and [0,1,0]. LDOS at $\mathbf{e}_d$ =
[0,0,1] and [0,1,0] is the same. (b, Media 7) point \textbf{b} for
$\mathbf{e}_d$ = [1,1,2], [-1,1,0] and [-1,-1,1]. (c, Media 8) point
\textbf{c} for $\mathbf{e}_d$ = [1,1,1] and [-1,1,0]. LDOS at
$\mathbf{e}_d$ = [-1,-1,2] is equal to that at $\mathbf{e}_d$ =
[-1,1,0]. (d, Media 9) point \textbf{d} for $\mathbf{e}_d$ =
[0.33,0.13,0], [-0.13,0,0.33] and [0,0,1].}
\end{figure}

In recent time-resolved experiments \cite{Lodahl04,NikolaevPRB07},
emission  from quantum dot light sources distributed at the internal
TiO$_2$-air interfaces of the inverse opals was investigated. To
analyze the experimental data, we have calculated the LDOS at
several symmetry-inequivalent positions on the TiO$_2$ shells: at
points \textbf{a}, \textbf{b}, \textbf{c} and \textbf{d} (see
Fig.~\ref{inverse-opal-unit-cell}) for three mutually orthogonal
orientations of the emitting dipole, where the first orientation is
chosen along the vector pointing from (0,0,0) toward the
corresponding point. Figures~\ref{LDOS-in-TiO2-on-shell}(a) through
\ref{LDOS-in-TiO2-on-shell}(d) (Media 6 through 9) show that at all
these positions the LDOS strongly varies with reduced frequency and
position, as expected, and also with orientation of the dipole. In
broad terms, for a dipole parallel to the interface the LDOS is near
1 at low frequency, increases to a peak enhancement of $2.5\times$
at $a/ \lambda = 1.22$ before strongly varying at high frequencies.
For reference, the frequency $a/ \lambda = 1.22$ is in the range
where a band gap is expected for more strongly interacting crystals.
The plots also reveal that for dipoles perpendicular to the
TiO$_2$-air interface, the LDOS is strongly inhibited (more than $10
\times$) over broad frequency ranges. For instance,
Figure~\ref{LDOS-in-TiO2-on-shell}(a) shows that the emission rate
for a dipole perpendicular to the interface is 16-fold inhibited to
a level of 0.06, with a maximum of 0.22 (5-fold inhibition) at $a/
\lambda = 1.22$. It is striking that the strong inhibition occurs
over a broad frequency range from 0.3 to 1.6, \emph{i.e.}, more than
two octaves in frequency. While the inhibition is not complete, the
bandwidth is much larger than the maximum bandwidth of $12\%$ for a
full 3D bandgap in inverse opals\cite{Koenderink03} and far exceeds
the bandwidth of the 2D gap for TE modes in membrane photonic
crystals, that allows a seven-fold
inhibition~\cite{Koenderink06,julsgaard08} over a 30\% bandwidth.

In the frequency range up to $a/\lambda = 1.1$, the dependence of
the LDOS on  frequency is quite similar at all the positions and
dipole orientations. This remarkable result agrees with the
experimental observation from Ref.~\cite{NikolaevPRB07}: the complex
decay curves of light sources in the inverse opals are described by
one and the same functional shape (log-normal) of the decay-rate
distribution for all reduced frequencies studied.

For the interpretation of the time-resolved experiments on ensembles of
emitters in the inverse-opal photonic crystals, the results
presented above mean that
\begin{enumerate}
\item the decay rate of an individual emitter is determined by its frequency,
position and also by the orientation of its transition dipole in the photonic crystal;
\item the measured spontaneous-emission decay depends on how the emitters are distributed in the
crystal;
\item even in the low-frequency regime, ensemble measurements will reveal non-exponentional decay curves;
\item the similar shape of the reduced-frequency dependence of the LDOS allows modeling of the non-exponentional  decay curves with a single type of decay-rate distributions. In other words, one distribution function can be successfully used to model the multi-exponentional decay curves measured from crystals with different lattice parameters.
\end{enumerate}

\section{LDOS in silicon inverse opals}\label{sec:Si}

\begin{figure}
\includegraphics[width=\figwidth]{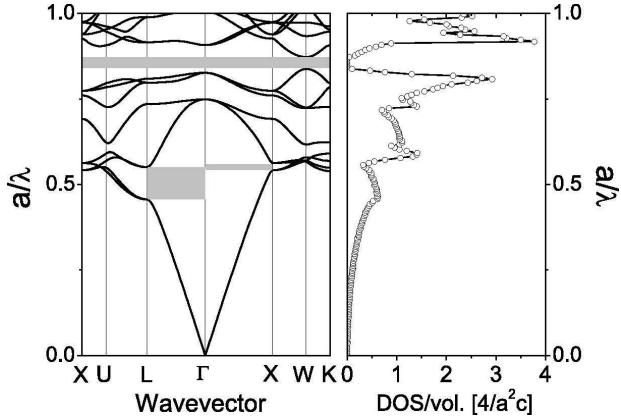}
  \caption{{\emph{Left}}: Photonic band structure (Media
10) for an inverse opal from silicon ($\epsilon$ = 11.9).
\emph{Right}: The total DOS in the Si inverse opal (Media 11). The
DOS is strongly depleted for frequencies near $\Gamma$L and
$\Gamma$X stopgaps (grey rectangles). A photonic band gap (grey bar)
occurs between bands 8 and 9, as also reflected in the vanishing
DOS.\label{DOS-in-Si}}
\end{figure}
A complete inhibtion of spontaneous emission may only be achieved in
photonic crystals  with a 3D photonic band gap. Therefore, there has
been much effort to fabricate inverse opals from silicon rather than
titania, since the higher index of silicon allows for a photonic
band gap~\cite{Busch98,Sozuer92}. For the calculation of the LDOS in
such Si inverse opals, the dielectric function
$\epsilon(\mathbf{r})$ was modeled similarly to that in the inverse
opals shown in Fig.~\ref{inverse-opal-unit-cell}. From SEM
observations we inferred the following structural
parameters~\cite{Blanco00,Vlasov01}: the outer radius of the
overlapping dielectric shells with $\epsilon = 11.9$ is about
1.15\emph{r} (where $r=0.25\sqrt{2}a$ is the air-sphere radius,
\emph{a} is the lattice parameter). The cylindrical windows
connecting neighboring air spheres have a radius of 0.2\emph{r}. The
larger outer radius and the smaller window size compared to the
TiO$_2$ inverse opals are commensurate with a higher volume fraction
of about 23\% Si~\cite{Vlasov01}.
\begin{figure}
\includegraphics[width=\figwidth]{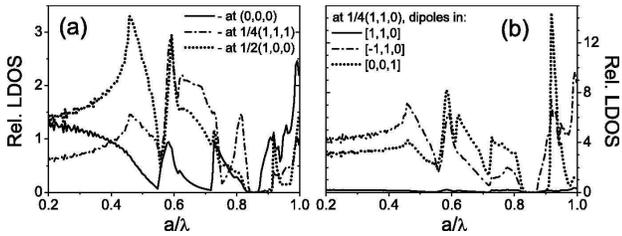}
  \caption{\label{LDOS-in-Si-1} Relative LDOS in a Si
inverse opal at: (a, Media 12) $\mathbf{r}$ = (0,0,0) [the center of
an air sphere, solid curve], $\mathbf{r}$ = $\frac{1}{4}(1,1,1)$
[among three air spheres, dash-dotted curve] and $\mathbf{r}$ =
($\frac{1}{2}$,0,0) [midway between two spheres along [1,0,0]
direction, dotted curve]; (b, Media 13) $\mathbf{r}$ =
$\frac{1}{4}(1,1,0)$ [in the window between two air spheres] projected
on [1,1,0], [-1,1,0] and [0,0,1] directions shown by solid,
dash-dotted and dotted curves, respectively.}
\end{figure}

The band structure and total DOS for this system are shown in
Fig.~\ref{DOS-in-Si} (Media 10 and Media 11).  The DOS is strongly
depleted in a pseudo-gap between the $2^{nd}$ and $3^{rd}$
bands~\cite{Ho90}, at frequencies near 0.55. Near frequency 0.85,
both the band structures and the DOS reveal a 3D photonic band gap
of relative width $\Delta \omega/\omega \approx 3~\%$ between the
$8^{th}$ and $9^{th}$ bands~\cite{Busch98}. Compared to the TiO$_2$
structure, both the lowest-order L-gap and the $8^{th}$ and $9^{th}$
bands are shifted to lower reduced frequencies. This shift is due to
the higher effective refractive index of the Si inverse opals
($n_{av} = 1.88$), on account of a higher index of the backbone and
a higher filling fraction.

Figure~\ref{LDOS-in-Si-1}(a) (Media 12) presents  the relative LDOS
at three high-symmetry positions in the unit cell \textbf{r} =
(0,0,0), $\frac{1}{4}(1,1,1)$ and $(\frac{1}{2},0,0)$. The LDOS
varies much more strongly with frequency than in TiO$_2$ inverse
opals, as a result of the larger dielectric contrast that leads to
strongly modified dispersion relations and Bloch mode profiles.
While the maxima up to 3.2 are not much higher than in TiO$_2$
inverse opals, the minima in the mode density are reduced and the
slopes are steeper. As expected, the LDOS is completely inhibited at
all positions in the frequency range of the band gap. Previously, it
has been suggested that the LDOS could be inhibited at salient
positions in the unit cell for frequencies outside a complete gap.
While Figure~\ref{LDOS-in-Si-1}(a) (Media 12) reveals that the mode
density is strongly reduced above the band gap, e.g., at
$\mathbf{r}=(\frac{1}{2},0,0)$, it is not truly inhibited. In fact,
in the course of our study, we have not encountered any "sweet
spots" where the LDOS is completely inhibited at frequencies outside
a 3D photonic band gap.

Figure~\ref{LDOS-in-Si-1}(b) (Media 13) shows the LDOS at a lower
symmetry position \textbf{r} = $\frac{1}{4}(1,1,0)$, in the window
between two nearest-neighbor air spheres. The mode density has been
calculated for three perpendicular dipole orientations.
Figure~\ref{LDOS-in-Si-1}(b) (Media 13) shows that the LDOS is
highly anisotropic since it differs for all three orientations: at
low frequencies ($a/ \lambda < 0.5$), the mode density is highest
for the $(-1,1,0)$ orientation, intermediate for the $(0,0,1)$
orientation, and lowest for the $(1,1,0)$ orientation. With
increasing frequency, the highest LDOS also changes to other
orientations, with the $(0,0,1)$ orientation having the highest LDOS
above the pseudo-gap and even a narrow peak at frequency 0.95.
Therefore, if it is possible to orient a quantum emitter with its
dipole parallel to $(0,0,1)$, this frequency range is conducive to
strong emission enhancement and perhaps even QED effects beyond
weak-coupling. Interestingly, these frequencies are slightly higher
than the upper edge of the band gap where strong coupling effects
have recently been discussed~\cite{Kristensen08}.

\section{Conclusions}
We have performed intensive calculations of the local density of
states in TiO$_2$ and Si inverse opals with experimentally relevant
structural parameters. Since conflicting and incorrect reports have
appeared on the LDOS in photonic crystals, we have set out to
validate our method of choice, i.e., the H-field plane-wave
expansion method. This validation relied on comparison to literature
results, on the explicit verification of required symmetries that
previous reports failed to satisfy, and on quantitative
considerations of resolution and accuracy. Results for each
structure are made available for other workers in the field, both as
benchmarks and for comparison with experimental data. With the help
of these computations we have obtained quantitative insight in the
LDOS relevant for time-resolved ensemble fluorescence measurements
on photonic crystals, such as obtained in recent experimental work
\cite{NikolaevPRB07}. The results of our numerical calculations
reveal a surprisingly strong dependence of the LDOS on the
orientation of the emitting dipoles.

\section*{Acknowledgments}
We thank Dries van Oosten for careful reading of the manuscript.
This work is part of the research program of the ``Stichting voor
Fundamenteel Onderzoek der Materie (FOM),'' which is financially
supported by the ``Nederlandse Organisatie voor Wetenschappelijk
Onderzoek (NWO)''. AFK and WLV were supported by VENI and VICI
fellowships funded by NWO. WLV also acknowledges funding by STW/NanoNed.

\end{document}